\documentclass[pdflatex,sn-mathphys-num]{sn-jnl}

\usepackage{graphicx}%
\usepackage{multirow}%
\usepackage{amsmath,amssymb,amsfonts}%
\usepackage{amsthm}%
\usepackage{mathrsfs}%
\usepackage[title]{appendix}%
\usepackage{xcolor}%
\usepackage{textcomp}%
\usepackage{manyfoot}%
\usepackage{booktabs}%
\usepackage{algorithm}%
\usepackage{algorithmicx}%
\usepackage{algpseudocode}%
\usepackage{listings}%
\usepackage{subcaption}

\theoremstyle{thmstyleone}%
\theoremstyle{thmstyletwo}%

\theoremstyle{thmstylethree}%

\begin{document}

\title[Article Title]{BoSS: Beyond-Semantic Speech}


\author[]{\fnm{Qing} \sur{Wang}}
\author[]{\fnm{Zehan} \sur{Li}}
\author[]{\fnm{Hang} \sur{Lv}}
\author[]{\fnm{Hongjie} \sur{Chen}}
\author[]{\fnm{Yaodong} \sur{Song}}
\author[]{\fnm{Jian} \sur{Kang}}
\author[]{\fnm{Jie} \sur{Lian}}
\author[]{\fnm{Jie} \sur{Li}}
\author[]{\fnm{Yongxiang} \sur{Li}}
\author[]{\fnm{Zhongjiang} \sur{He}}
\author*[ ]{\fnm{Xuelong} \sur{Li}}\email{xuelong\_li@ieee.org}  

\affil[]{Institute of Artificial Intelligence (TeleAI), China Telecom, China}




\abstract{
Human communication involves more than explicit semantics, with implicit signals and contextual cues playing a critical role in shaping meaning. However, modern speech technologies, such as Automatic Speech Recognition (ASR) and Text-to-Speech (TTS) often fail to capture these beyond-semantic dimensions. 
To better characterize and benchmark the progression of speech intelligence, we introduce \textbf{Spoken Interaction System Capability Levels (L1–L5)}, a hierarchical framework illustrated the evolution of spoken dialogue systems from basic command recognition to human-like social interaction. 
To support these advanced capabilities, we propose \textbf{Beyond-Semantic Speech (BoSS)}, which refers to the set of information in speech communication that encompasses but transcends explicit semantics. It conveys emotions, contexts, and modifies or extends meanings through multidimensional features such as affective cues, contextual dynamics, and implicit semantics, thereby enhancing the understanding of communicative intentions and scenarios.
We present a formalized framework for BoSS, leveraging cognitive relevance theories and machine learning models to analyze temporal and contextual speech dynamics. 
We evaluate BoSS-related attributes across five different dimensions, reveals that current spoken language models (SLMs) are hard to fully interpret beyond-semantic signals. These findings highlight the need for advancing BoSS research to enable richer, more context-aware human-machine communication.
}

\keywords{Beyond-Semantic Speech, Spoken Interaction System, Capability Levels, Paralinguistics}



\maketitle

\section{Introduction}\label{sec1}

In human communication, speech conveys not only explicit meanings, but also a wealth of implicit and contextual information. Beyond the words themselves, subtle vocal and contextual hints play a crucial role in shaping the listener’s understanding and response. 
Imagine a team meeting where a colleague subtly coughs while you discuss a sensitive topic. This cough, while not having explicit words, might serve as a deliberate signal: \textit{“Let’s move on to a different subject”}.
Similarly, in a friendly conversation, a sarcastic comment like “\textit{Great job!}” with a playful tone can invert its literal meaning, implying criticism rather than praise. 
Both scenarios rely on subtle vocal and contextual cues to convey meanings that go beyond the words themselves. 
This observation highlights the complexities of human communication, where subtle signals beyond the explicit semantics of words carry rich and nuanced meanings.

\begin{figure}
    \centering
    \includegraphics[width=1\linewidth]{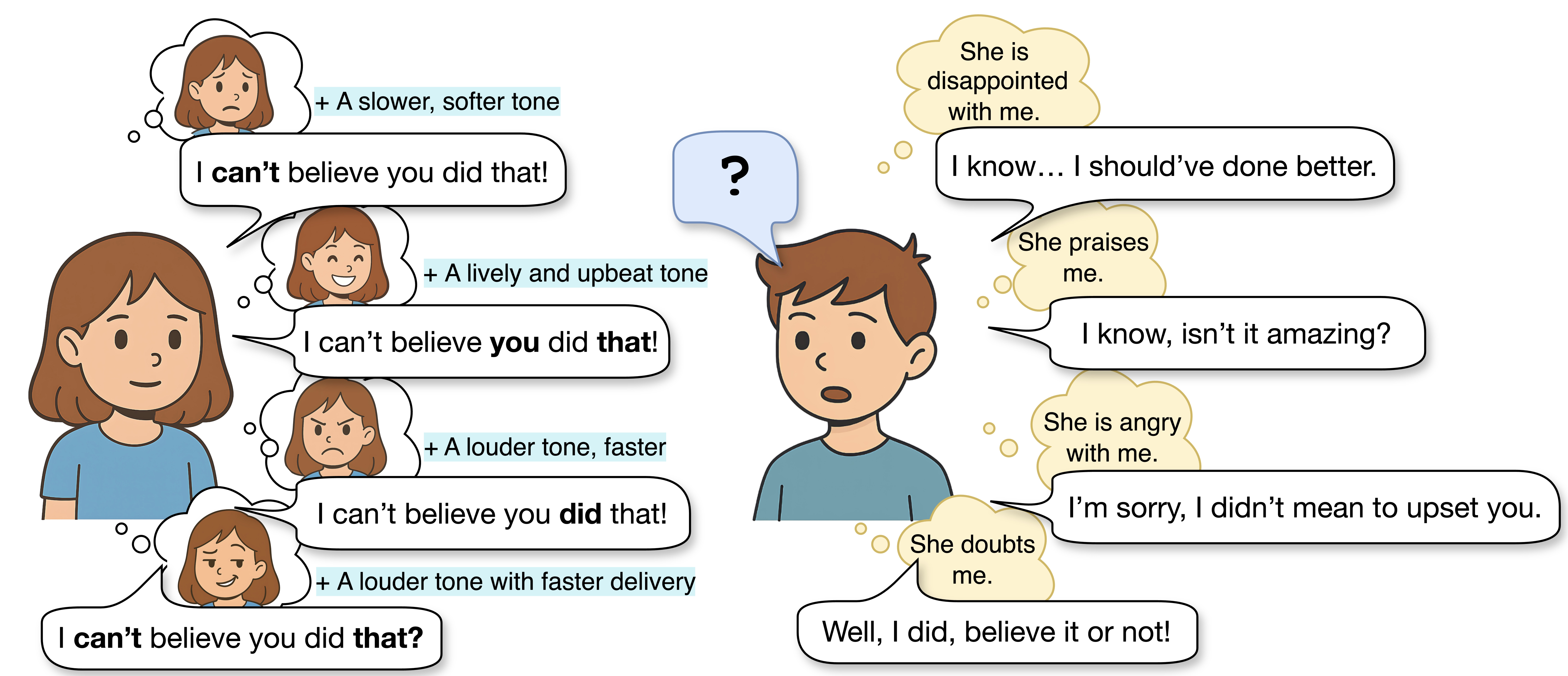}
    \caption{Schematic of How Vocal Cues and Expression Influence the Interpretation of Speech}
    \label{fig:overall}
\end{figure}

Consider the phrase “\textit{I can’t believe you did that.}” as illustrated in Figure~\ref{fig:overall}. Depending on how it is expressed, the same phrase can have entirely different meanings. A slow pace, downward tone, and emphasis on “can’t” might signal disappointment, evoking feelings of guilt or remorse in the listener. A bright tone, quick pace, and emphasis on “you” and “that” could convey positive amazement, leaving the listener feeling appreciated. In contrast, a loud and fast delivery with emphasis on “did” might suggest frustration or blame, prompting a defensive response. Alternatively, a questioning tone with rising intonation at the end emphasizes disbelief, encouraging the listener to explain or justify their actions. These examples demonstrate how intonation, vocal tone, and emphasis shape emotional and cognitive interpretations, underscoring the limitations of relying solely on explicit semantic content and shaping listener's emotional and cognitive responses in multiple ways.

Despite the richness of human speech, contemporary speech technologies, such as Automatic Speech Recognition (ASR) and Text-to-Speech (TTS), primarily focus on extracting and synthesizing explicit semantics, the literal meaning of words. However, real-world communication is far more intricate. Humans effortlessly integrate implicit signals, such as intonation, emotional cues, and pauses, as well as contextual signals, including conversational history and environmental sounds, to infer nuanced meanings and intentions. However, current speech systems struggle to replicate this human capacity. This limitation raises critical questions: \textit{How can machines identify and interpret the non-verbal and contextual cues embedded in speech? What are the inherent multi-dimensional features of speech that enable it to convey layered meanings? How can these features advance machine understanding of human communication?}

\begin{figure}
    \centering
    \includegraphics[width=0.56\linewidth]{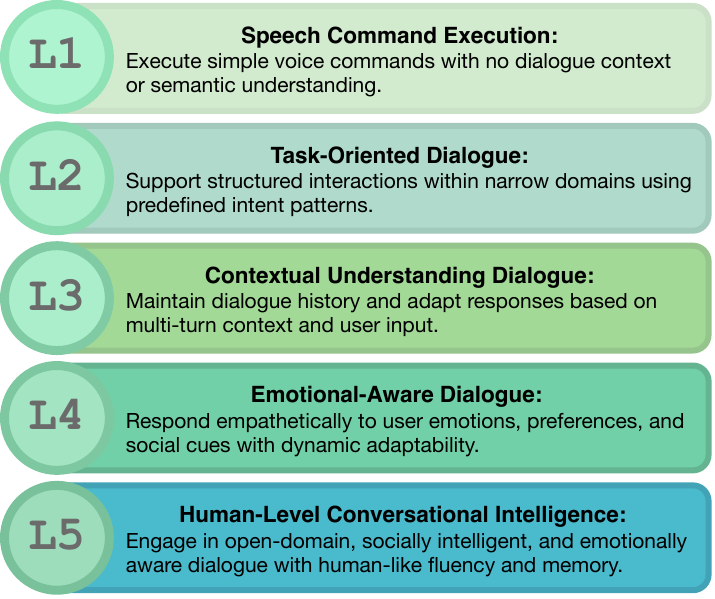}
    \caption{Spoken Interaction System Capability Levels (L1–L5)}
    \label{fig:level}
\end{figure}

To address these questions and assess the evolving intelligence of speech-based systems,  
we introduce a hierarchical framework: \textbf{Spoken Interaction System Capability Levels (L1–L5)}, as shown in Figure~\ref{fig:level}. 
Inspired by the classifications in autonomous driving~\cite{on2021taxonomy}, this framework categorizes systems from basic command recognition to human-level social interaction, this L1–L5 framework reflects increasing degrees of linguistic understanding, context awareness, emotional intelligence, and interactional flexibility. 
As systems progress towards Levels 4 and 5, they are expected to engage in complex, multi-domain conversations, interpret speaker attitudes and intentions, and exhibit empathy or social awareness. These expectations place growing demands on the system’s ability to interpret information beyond surface-level semantics. These capabilities require moving beyond literal understanding to include the broader communicative dimensions of speech.

To support this goal, in this study, we propose \textbf{Beyond-Semantic Speech (BoSS)}, designed to explore the multi-dimensional characteristics of speech that extend beyond explicit semantics. BoSS enables machines to understand the subtle implications and nuanced meanings embedded in human speech. Beyond-Semantic Speech refers to the set of information in speech communication that encompasses but transcends explicit semantics. It conveys emotions, shapes contexts, and modifies or extends meanings through multidimensional features such as vocal characteristics, dynamic contexts, and implicit semantics, thereby enhancing the understanding of communicative intentions and scenarios.
BoSS integrates four primary dimensions:
\begin{itemize}
    \item \textbf{Explicit Semantics}: The foundational layer comprising \textbf{lexical}, \textbf{syntactic}, and \textbf{logical} structures that convey direct meaning through factual statements or clear commands.
    \item \textbf{Affective Cues}: The emotional subtext carried through \textbf{vocal characteristics} (pitch, rhythm, volume) and \textbf{non-verbal vocalizations} (laughter, sighs), as well as \textbf{emotional expressions} that reveal the speaker's internal state and attitude.
    \item \textbf{Contextual Dynamics}: The environmental and interactional context including \textbf{background sounds}, \textbf{acoustic properties} of space, \textbf{conversational rhythm} (pauses, turn-taking), and shared \textbf{interaction history} that shape meaning.
    \item \textbf{Implicit Semantics}: The inferred meanings include \textbf{semantic modifications} (where tone alters literal meaning), \textbf{indirect intent} (sarcasm, irony), and \textbf{identity cues} that allow listeners to deduce social roles and relationships.
\end{itemize}

These BoSS dimensions contribute not only to \textbf{speech comprehension} but also to \textbf{speech generation}. For comprehension, they enable systems to reliably perceive and interpret beyond-semantic signals, thus allowing systems to understand the speaker’s true intent beyond literal meaning. 
For generation, BoSS attributes can be directly used to synthesize speech with specific emotional or contextual characteristics, rather than merely relying on semantic input. This allows systems to produce speech that is more natural, expressive, and context-appropriate, better capturing the richness and subtlety of human communication.

This study also evaluates several BoSS-related attributes to examine their effects in enhancing machine understanding of speech communication. By focusing on these dimensions, this work establishes the foundation for integrating multi-dimensional features of speech into intelligent systems, enabling them to bridge the gap between explicit and implicit communication and paving the way to more human-like interactions.

\section{Survey of Related Work}
In this section, the development related to Beyond-Semantic Speech (BoSS) in the field of phonetics will be introduced, including its early stage, developmental phase, and the era of large audio language models.

\subsection{Early Stage of Beyond-Semantic Speech}

The exploration of BoSS is originated from the study of paralinguistic information~\cite{grice1957meaning, mehrabian1971silent, crystal1975paralinguistics}, which aims to uncover the meanings embedded in speech that extend beyond explicit semantics. 
Early research has laid the conceptual groundwork for identifying and analyzing these non-explicit elements. 

In 1958, Trager~\cite{trager1958paralanguage} introduced Paralanguage, highlighting non-verbal features like intonation, pauses, and speech rate that convey emotions, attitudes, and social roles. Building on this, Crystal~\cite{crystal1969prosodic} formalized the concept of Paralinguistics and initiated efforts to systematically categorize such features. These foundational studies emphasized how these elements complement or modify explicit semantics in speech.

Theoretical advances followed in the 1970s and 1980s, providing cognitive frameworks to understand implicit meaning. Grice’s Cooperative Principle~\cite{grice1975logic} demonstrated how adherence to or violations of conversational maxims (e.g., relevance or quantity) convey hidden meanings such as sarcasm or humor. 

\subsection{Developmental Phase of Beyond-Semantic Speech}

During the 1980s and 1990s, the study of speech meaning~\cite{pierrehumbert1980phonology, cutler1997prosody} moved from descriptive and theoretical paradigms to cognitive and computational approaches. Sperber and Wilson’s Relevance Theory~\cite{sperber1986relevance} provided a cognitive model for understanding speech, emphasizing the trade-off between cognitive effort and effect. They posited that effective communication minimizes cognitive processing while maximizing contextual understanding, laying the groundwork for the computational modeling of implicit meaning in speech. At the core of Relevance Theory lies the principle of optimal relevance, which suggests that humans aim to maximize the cognitive effects of a message while minimizing the effort required to process it. The formula for relevance is expressed as follows.
\begin{equation}
    R = \frac{E}{P},
\end{equation}
where, $R$ is relevance, representing the importance of a message or signal in the current context. $E$ is cognitive effect, quantifying the contribution of a message to the communicative goal or cognitive state (e.g., introducing new knowledge, reinforcing beliefs, or correcting misconceptions) and $P$ is processing effort, representing the cognitive cost of interpreting and understanding the message.
In this framework, higher relevance corresponds to scenarios where the cognitive effect $E$ significantly outweighs the processing effort $P$. In contrast, lower relevance indicates situations where the effort required to understand the information is almost equal to or greater than its cognitive benefit.

As computational tools became increasingly accessible, researchers began exploring ways to automate the extraction and interpretation of speech features. Picard’s introduction of “Affective Computing”~\cite{picard2000affective} marked a pivotal moment by modeling the emotional elements of speech. Although initially centered on emotional recognition, affective computing inspired a broader exploration of implicit speech features, including attitudes and social cues.

In 2010, Schuller and colleagues~\cite{schuller2013paralinguistics} formalized “Computational Paralinguistics”, transitioning the field into a data-driven era. This approach integrated machine learning techniques to analyze paralinguistic features such as speaker identity, emotion, and conversational attitude. The shift from descriptive studies to computational frameworks enabled large-scale analyses of diverse speech datasets, providing deeper insights into the nuances of non-semantic information in speech.

\subsection{Beyond-Semantic Speech in Large Audio-Language Model}
Speech represents the most natural and fundamental medium for human interaction. With ongoing technological advancements, there is a growing aspiration for human-computer interaction to approximate the fluidity and subtlety of human-human communication, ultimately progressing toward a speech-level Turing test. The release of Whisper~\cite{Radford2022RobustSR} has marked a critical milestone, signaling the advent of the era of Large Audio-Language Models (LALMs). As research on large language models continues to evolve, increasing attention is being directed toward the anthropomorphic dimensions of human-computer interaction. In particular, emerging studies have begun to examine how beyond-semantic speech factors, such as affective cues, influence the naturalness and effectiveness of speech-based interaction.

\subsubsection{Generation Large Audio-Language Model}
As a classic generative task, text-to-speech (TTS) and voice conversion (VC) have made remarkable progress in the era of large-scale models. Beyond maintaining high accuracy in speech generation, increasing attention has been directed toward more beyond-semantic fine-grained control over speech characteristics, such as emotion and prosody modulation. These efforts have already yielded promising initial results. 

Representative studies in this direction include the following. 
Make-a-Voice~\cite{Huang202305MakeAVoiceUV} introduces discrete audio representations that include a variety of acoustic conditions (e.g., speaker, emotion, prosody, and style) to generate the speech. Meanwhile, Fish‑Speech~\cite{Liao202411FishSpeechLL} unifies LLM-driven discrete representations to support multilingual and multi-speaker synthesis. Then, UniAudio~\cite{Yang202310UniAudioAA} covers 11 audio generation tasks, and timbre, emotion, prosody are considered during generation. And, AnyGPT~\cite{Zhan202402AnyGPTUM} extracts the timbre and emotion under the prompt for generation.
Furthermore, MatchaTTS~\cite{Mehta202309MatchaTTSAF} innovatively employs optimal-transport conditional flow matching to achieve prosody-controlled rapid synthesis. Other architectures, such as models guided by synthetic prosody annotations, integrate emotion labels directly into inputs to guide expressive control, while systems like Seed‑TTS~\cite{Anastassiou202406SeedTTSAF} and F5‑TTS~\cite{Chen202410F5TTSAF} leverage style embeddings and flow matching to flexiblely adjust speaker characteristics. In addition, Parler-TTS~\cite{Lyth202402NaturalLG} uses natural language to control emotion, stress, and style. Recently, EmergentTTS-Eval has been introduced as a benchmark designed to more effectively evaluate complex prosodic, expressive, and linguistic challenges in TTS tasks.
In addition, significant progress has also been made in the field of music~\cite{Agostinelli202301MusicLMGM} and audio generation~\cite{Yang202207DiffsoundDD}.

\subsubsection{Understanding Large Audio-Language Model}
In contrast to generation tasks, which need fine-grained and beyond-semantic control, the understanding tasks that analyze the input comprehensively are also critical. This analysis should encompass both accurate textual content and beyond-semantic information, enabling a more precise and faithful expression of thought.

In Large Audio-Language Models, a high-quality audio encoder is vital. Today, whisper-based speech encoders have been widely adopted for extracting semantic information from input audio. Meanwhile, effective encoding of beyond-semantic information for a more comprehensive understanding of the input audio is equally essential. To this end, researchers have proposed models such as Emotion2Vec~\cite{Ma202312emotion2vecSP} which captures emotional cues from speech; SpeechTokenizer~\cite{Zhang202308SpeechTokenizerUS} which employs residual vector quantization (RVQ) to jointly model both semantic and acoustic information; and Salmonn~\cite{Tang202310SALMONNTG} leverages a Q-Former architecture to explicitly model acoustic features independent of semantic content.
 
Building upon the appropriate semantic and beyond-semantic audio encoders, a diverse range of large models have been developed for enabling more advanced and diverse audio understanding tasks. For example, AudioGPT~\cite{Huang202304AudioGPTUA} supports tasks such as sound event detection; VideoChat~\cite{Li202305VideoChatCV} enhances question-answering performance by incorporating emotion analysis from video content; and AudioPaLM~\cite{Rubenstein202306AudioPaLMAL} not only addresses traditional ASR / TTS tasks, but also preserves paralinguistic information in speech-to-speech translation (S2ST) task. Joint Audio and Speech Understanding~\cite{Gong202309JointAA} further explores the paralinguistic and audio processing capabilities of Whisper-based encoders. LauraGPT~\cite{Wang202310LauraGPTLA} and WavLLM~\cite{Hu202403WavLLMTR} extend this line of research to emotion recognition. GAMA~\cite{Ghosh202406GAMAAL} integrates both non-verbal speech and non-speech sound information to enhance reasoning capabilities. Audio Flamingo~\cite{Kong202402AudioFA} and Audio Flamingo 2~\cite{Ghosh2025AudioF2} enable long-form audio understanding with improved inference of non-verbal information. Most recently, Audio-Reasoner~\cite{Xie2025AudioReasonerIR} introduces an explicit “Planning-Captioning-Reasoning-Summary” chain-of-thought (CoT) framework to improve the integration and utilization of both semantic and beyond-semantic information.

Obviously, the development of LALMs is inseparable from the availability of high-quality benchmarks, which play a crucial role in driving progress. Initiatives such as AudioBench~\cite{Wang202406AudioBenchAU}, Dynamic-SUPERB~\cite{Huang202311DynamicSuperbTA}, and AIR-Bench~\cite{Yang202402AIRBenchBL} have been introduced to provide diverse tasks that aim to comprehensively cover various aspects of audio understanding. However, there remains substantial room for improvement in the evaluation and coverage of beyond-semantic information within these benchmarks.

\subsubsection{Dialogue Large Audio-Language Model}
Compared to input-side understanding tasks and output-side generation tasks, dialogue systems place significantly higher demands on the comprehensive capabilities of models. Such systems must accurately analyze and interpret both semantic and beyond-semantic information over long conversational contexts. Furthermore, they require the ability to generate distinct, contextually appropriate responses to different expressions, enabling human-machine interaction that is both explicit and natural.

Recently, dialogue-oriented LALMs that, to some extent, leverage beyond-semantic information have begun to emerge as an active area of exploration. For instance, SpeechFormer++~\cite{Chen202302SpeechFormerAH} strengthens the processing and response capabilities related to emotional states, depression, and neuro-cognitive disorders within conversational settings. Qwen-Audio~\cite{Chu2023QwenAudioAU} introduces a label-based approach to incorporate emotion perception and non-verbal events into dialogue modeling, while Qwen-Audio 2~\cite{Chu2024Qwen2AudioTR} further advances emotion and semantic analysis and response generation in natural dialogue scenarios. ParalinGPT~\cite{Lin2023ParalinguisticsEnhancedLL} integrates emotion and prosody information to enhance dialogue expressiveness. Similarly, E-Chat~\cite{Xue2023EChatES} incorporates sentiment information to improve conversational effectiveness. LLaMA-Omni~\cite{Fang2024LLaMAOmniSS} adopts a dual-modal input of speech and text, focusing on modeling stylistic dimensions of dialogue such as clarity and naturalness.

\section{Spoken Interaction System Capability Levels}
Inspired by the hierarchical levels of autonomous driving (Level 0–5), we introduce a Spoken Interaction System Capability Framework (Level 1–5) to systematically categorize and benchmark the progression of machine intelligence in speech-based human–machine interaction, shown in Figure~\ref{fig:level}. Each level reflects increasing capability in natural language understanding, contextual awareness, and beyond-semantic comprehension.
\begin{itemize}
    \item \textbf{Level 1 – Speech Command Execution:} 
    The system supports basic keyword-based or single-turn voice commands. It functions similarly to a remote control, with no contextual memory or semantic understanding beyond predefined instructions.
    \item \textbf{Level 2 – Task-Oriented Dialogue:} 
    The system is able to engage in structured dialogues within a limited domain, supporting task completion such as travel bookings or service inquiries. Its understanding of user input is based on predefined intents and shallow logic, with minimal flexibility or adaptation beyond fixed scenarios.
    \item \textbf{Level 3 – Contextual Understanding Dialogue:} 
    The system is capable of maintaining conversational context over multiple turns. It can resolve references, track dialogue state, and adapt responses based on dialogue history. However, its understanding is still mostly task-driven, with limited sensitivity to nuanced speaker signals.
    \item \textbf{Level 4 – Emotion-Aware Dialogue:} 
    The system begins to integrate user modeling and affective perception, allowing it to adjust its tone, phrasing, and conversational strategy based on detected emotional states, speaking style, or user preferences. This stage introduces more flexible dialogue management and improves the user experience in more natural and personalized interactions.
    \item \textbf{Level 5 – Human-Level Conversational Intelligence Dialogue:} 
    The system demonstrates advanced conversational capabilities across open domains. It can handle nuanced linguistic signals, perceive implicit intentions, and respond to emotion and social context with appropriate verbal behavior. In addition to context tracking and emotional perception, the system exhibits regulatory mechanisms such as self-adjustment in response to conflict or ambiguity, integration of diverse viewpoints, and consistent interaction over an extended dialogue history. This level marks a shift from functional assistance to socially aware and emotionally aligned communication.
\end{itemize}
This five-level framework provides a reference standard for analyzing and benchmarking the capabilities of spoken interaction systems. Each level reflects increasing competence in language understanding, contextual integration, and interaction strategy. Particularly, higher levels (L4–L5) emphasize the system’s ability to process and respond to beyond-semantic signals, such as emotional tone, speaker intent, and situational context, which are essential to build more natural and human-aligned dialogue systems.

However, achieving such advanced capabilities hinges on the system’s ability to interpret information beyond explicit semantics. This necessitates a more comprehensive understanding of speech that integrates emotional expression, dynamic context, and implicit meaning—dimensions we collectively define as Beyond-Semantic Speech (BoSS). 

\section{Problem Formulation}

\subsection{Beyond-Semantic Speech: Optimal Meaning Hypothesis}
The goal of the BoSS framework is to determine the optimal meaning hypothesis $H_t^*$ at each time step $t$, which represents the most plausible interpretation of the speaker’s intent, emotion, or underlying meaning. This hypothesis is selected by maximizing the ratio of cognitive effects $\mathcal{E}_H$ to processing effort $\mathcal{P}_H$. Mathematically, $H_t^*$ is defined as:
\begin{equation}\label{eq:overall}
H^*_t = \text{argmax}_{H \in \mathcal{H}} \left( \frac{\mathcal{E}_H}{\mathcal{P}_H} \right).
\end{equation}
Here, $\mathcal{E}_H$ quantifies the informativeness of a hypothesis $H$, measuring how much it contributes to understanding the communication. In contrast, $\mathcal{P}_H$ measures the cognitive and computational cost required to interpret $H$. 

In Equation~\ref{eq:overall}, $\mathcal{E}_H=\mathcal{E}(H, O_t, \mathcal{C}_t)$ and $\mathcal{P}_H=\mathcal{P}(H, O_t, \mathcal{C}_t)$.
Here, $O_t$ is the observation vector containing features derived from speech, while $\mathcal{C}_t$ represents the external context, such as conversational history and environmental cues. This framework ensures that the system identifies meanings that are both relevant and computationally efficient to process.

The observation vector $O_t$ encapsulates critical aspects of the speech signal, integrating explicit semantic, emotional, contextual, and implicit features. It is represented as:
\begin{equation}\label{eq3}
O_t = [V_{L,t}, V_{AC,t}, V_{CD,t}, V_{IS,t}],
\end{equation}
where, $V_{L,t}$ is the explicit semantics latent vector, representing the lexical content extracted from Automatic Speech Recognition (ASR) systems. It captures the literal meaning of speech. 
$V_{AC,t}$ is the affective cues latent vector, encoding emotional and prosodic features, such as pitch, tone, and intensity, which are indicative of emotions like happiness, sadness, or anger.
$V_{CD,t}$, the contextual dynamics latent vector, modeling interaction patterns, environmental sounds, and pauses to infer contextual signals. For example, a long pause might suggest hesitation or confusion.
$V_{IS,t}$, the implicit semantics latent vector, capturing deeper meanings like sarcasm, humor, or emphasis, often inferred through the interplay of lexical content and non-verbal signals.

This multi-dimensional vector allows the system to holistically understand the nuances of speech.
The external context vector $\mathcal{C}_t$ plays a crucial role in disambiguating meanings by providing information beyond the immediate speech signal. It is made up of the following components:
\begin{equation}
\mathcal{C}_t = [C_{\text{Hist},t}, C_{\text{Env},t}, C_{\text{Char},t}, C_{\text{Task},t}],
\end{equation}
where $C_{\text{Hist},t}$ captures conversational history, such as previous dialogue turns, which provides continuity and helps resolve ambiguous references.
$C_{\text{Env},t}$ models environmental factors, like background noise or channel type (e.g., phone call vs. face-to-face), which may influence interpretation.
$C_{\text{Char},t}$ incorporates speaker and listener characteristics, including speaker identity, emotional state, or personal preferences, which affect the way speech is perceived.
$C_{\text{Task},t}$ encodes task-specific information or domain knowledge, helping to understand in specialized contexts, such as technical discussions or customer support.

By integrating $\mathcal{C}_t$ with $O_t$, the system can better interpret speech in dynamic and complex scenarios.
The relevance of hypothesis $H$ is controlled by two key metrics: cognitive effects $\mathcal{E}$ and processing effort $\mathcal{P}$. These metrics are defined as follows:
\begin{equation}
\mathcal{E}(H, O_t, \mathcal{C}_t) = \text{NN}_{\mathcal{E}}([H_{\text{emb}}, O_t, \mathcal{C}_t]),
\end{equation}
\begin{equation}
\mathcal{P}(H, O_t, \mathcal{C}_t) = \text{NN}_{\mathcal{P}}([H_{\text{emb}}, O_t, \mathcal{C}_t]),
\end{equation}
where $\mathcal{E}$ measures how much new, relevant knowledge the hypothesis $H$ brings to the interpretation of the speech signal. For example, identifying sarcasm from a mismatch between tone and lexical content generates a high cognitive effect. $\mathcal{P}$ quantifies the effort needed to evaluate the hypothesis $H$. Complex signals, conflicting cues, or ambiguous contexts increase $\mathcal{P}$.
These metrics are implemented as neural networks ($\text{NN}_{\mathcal{E}}$ and $\text{NN}_{\mathcal{P}}$) that learn to evaluate relevance based on multimodal inputs.

The BoSS framework leverages HMMs to model the temporal dynamics of speech signals. The emission probability, which links hidden states $H_t$ to observations $O_t$, is derived as:
\begin{equation}
P(O_t | H_t = H_j) = \frac{\exp\left(\frac{\mathcal{E}(H_j, O_t, \mathcal{C}_t)}{\mathcal{P}(H_j, O_t, \mathcal{C}t)}\right)}{\sum{H_k \in \mathcal{H}} \exp\left(\frac{\mathcal{E}(H_k, O_t, \mathcal{C}_t)}{\mathcal{P}(H_k, O_t, \mathcal{C}_t)}\right)}.
\end{equation}
This formulation integrates the relevance computation directly into the probabilistic framework, enabling the system to select the most likely sequence of hidden states using algorithms like Viterbi decoding.

The overall objective of the BoSS framework is to learn the parameters of the various encoders and neural networks such that the system can accurately infer the sequence of beyond-semantic states. This is implicitly achieved by maximizing the likelihood of the observed sequences given the inferred hidden states, effectively optimizing the relevance score for correct interpretations while minimizing it for incorrect ones. The training process would involve optimizing the HMM parameters (transitions) and the parameters of $\text{NN}_{\mathcal{E}}(\cdot)$ and $\text{NN}_{\mathcal{P}}(\cdot)$	(for emissions), leveraging labeled datasets of beyond-semantic meanings.

\subsection{End-to-end Spoken Language Models}
More specifically, taking end-to-end Spoken Language Models (SLMs) as an example, we can obtain the Optimal Meaning Hypothesis by maximizing mutual information.
SLM typically consists of three main components: a speech encoder, a large language model (LLM), and a text-to-speech (TTS) decoder. 

For any continuous speech signal \(U \in [U_1, U_2, ..., U_T] \), the speech encoder extracts either a continuous representation or a sequence of discrete unit \(Z_w(V_L, V_{AC}, V_{CD}, V_{IS})\), abbreviated as \(Z_w\). 
This process can be succinctly represented by a nonlinear mapping \(F_w(.\ ; \theta_w)\), such that \(F_w(U;\theta_w) \to Z_w\). The LLM processes both the semantic and acoustic information contained in the input, and generates a corresponding response embedding \(Z_L(V_L, V_{AC}, V_{CD}, V_{IS})\), abbreviated as \(Z_L\). We abstract the operation of the LLM as a nonlinear mapping \(F_L(.\ ; \theta_L)\), such that \(F_L(Z_w;\theta_w) \to Z_L\). The TTS decoder transforms the output of the LLM into the final speech waveform \(Y\). 

In an end-to-end spoken language model (SLM), in order to preserve more information, the input to the TTS decoder is typically an embedding or a sequence of discrete speech units, rather than text. The operation of TTS decoder can be summarized as \(F_v(.\ ; \theta_v)\), \(F_v(Z_L;\theta_v) \to Y\). \(F_v\) is expected to model all the attribute information contained in \(Z_L\). Then, the mutual information between \(U\) and \(Z_L\) has an upper bound \(E\left \{ D_{KL} [P(Z_L|Z_w) || V(Z_L)] \right \} \), where \(V(.)\) is a theoretical distribution from which the TTS decoder can fully reconstruct both the semantic and beyond-semantic information contained. The difference between the probability distributions of \(P(Z_L|Z_w)\) and \(V(Z_L)\) can be measured using the following Kullback-Leibler (KL) divergence calculation formula.

{\small
\begin{equation}\label{eq:slm}
\begin{aligned}
    D_{\mathrm{KL}}\left[P(Z_L|Z_w)\,\|\, V(Z_L)\right] 
    &= \sum_{Z_L} P(Z_L|Z_w) \log \frac{P(Z_L|Z_w)}{V(Z_L)} \\
    &= \sum_{Z_L} P(Z_L|Z_w) \log\left( \frac{P(Z_L|Z_w)}{P(Z_L)} \right)
    + \sum_{Z_L} P(Z_L|Z_w) \log\left( \frac{P(Z_L)}{V(Z_L)} \right).
\end{aligned}
\end{equation}
}

Ideally, the speech encoder compresses \(U\) without losing semantic or acoustic information, ensuring that the LLM yields consistent outputs when processing either \(U\) or \(Z_w\), which can be expressed as \(P(Z_L|U) = P(Z_L|Z_w)\). Taking the expectation over \(U\) on both sides of Equation~\ref{eq:slm}, we obtain the following:

{\small
\begin{equation}
\begin{aligned}
E_U\left\{ D_{\mathrm{KL}}\left[P(Z_L|Z_w)\,\|\, V(Z_L)\right] \right\}
&= E\left\{ D_{\mathrm{KL}\left[P(Z_L|U)\,\|\, P(Z_L)\right]} \right\} 
+ E_U\left\{ \sum_{Z_L} P(Z_L|U)\log\left( \frac{P(Z_L)}{V(Z_L)} \right) \right\} \\
&= I(Z_L; U) + D_{\mathrm{KL}}\left[ P(Z_L)\,\|\, V(Z_L) \right].
\end{aligned}
\end{equation}
}

From the non-negativity of the KL divergence, it follows that \(D_\mathrm{KL}\left [ P(Z_L)||V(Z_L) \right ]\ge 0\), then we have \(I(Z_L;U)\le E\left \{ D_{KL}[P(Z_L|Z_w)|| V(Z_L)] \right \}\). The maximum upper bound can be achieved if and only if the distributions \(P(Z_L)\) and \(V(Z_L)\)are identical. Similarly, by applying the Barber–Agakov decomposition, we can obtain a variational lower bound \(E\left \{ log\ q_\phi(Z_L|U)-logP(Z_L) \right \} \). LLM is aimed to find the mapping \(F_L\) that approximates this variational distribution. Under these circumstances, \(\phi=\theta\).

\section{Experiments and Results}

In this study, we conduct an initial evaluation of several dimensions of Beyond-Semantic Speech (BoSS) attributes, including Chinese dialect comprehension and generation, context memory, emotion perception and response, age perception and response, and non-verbal information. These attributes are chosen as they represent key dimensions of BoSS in Equation~\ref{eq3}, encompassing both explicit and implicit factors that influence natural and human-like speech understanding.

To facilitate the evaluation, we construct several evaluation datasets aiming at these dimensions and select open source speech language models (SLM)~\cite{zeng2024glm,Zhan202402AnyGPTUM,yao2024minicpm,li2025baichuan,sg2preview, wang2024freeze, xu2025qwen2, ding2025kimi, Fang2024LLaMAOmniSS} that support Chinese speech input and output as test models. The evaluations aim to assess how effectively these models integrate BoSS-related attributes into their responses and interactions.

\subsection{Chinese Dialect Comprehension and Generation}
To assess how well the system handles phonetic variation and lexical diversity, we evaluate its ability to comprehend and follow conversations in different Chinese dialects. This task tests both explicit semantics (lexical understanding) and affective cues (especially vocal characteristics like accent and pronunciation), revealing how well the model generalizes beyond standard speech.

\subsubsection{Dataset}
We construct two datasets, \textit{aqa-\{dialect\}} and \textit{chitchat-\{dialect\}}, to evaluate dialect comprehensive and generation performance of these models across five Chinese dialects: Cantonese, Henan dialect, Northeastern Mandarin, Shanghainese, and Sichuanese. Specially, \textit{aqa-\{dialect\}} is designed as a dialectal audio question-answer (AQA) dataset to evaluate models’ ability to understand dialectal audio input. The expected output is a response in Mandarin text and the evaluation metric is accuracy. While, \textit{chitchat-\{dialect\}} serves as a dialectal chit-chat dataset, targeting models’ ability to understand and follow up on dialectal speech. The input is user speech in a dialect, and the expected output is a text response in the same dialect. Evaluation focuses on dialectal consistency and semantic appropriateness.

The user query texts in the \textit{aqa-\{dialect\}} evaluation set are sourced from \textit{chineseQuiz}, a Chinese cultural knowledge dataset, while those in the \textit{chitchat-\{dialect\}} set are manually selected from MagicData~\cite{yang2022open}. We apply a dialect translation model to rewrite these queries, adapting them to the lexical and syntactic characteristics of the target dialects. Subsequently, for each dialect, we use a dialectal speech synthesis system to generate audio by randomly selecting from 30 available prompt speakers.

To ensure the quality of the two datasets, each regional dialect audio sample is evaluated by ten native speakers of the corresponding dialect, who assess it based on the naturalness of pronunciation and the semantic coherence. 

\subsubsection{Results}
To assess dialectal comprehension, we adopt a string-matching strategy to determine answer correctness, foregoing conventional LLM-based scoring methods. This approach ensures the consistency of the evaluation and mitigates the ambiguity and boundary issues previously reported in ~\cite{cui2025voxeval}. 
For each AQA (Answer Questioning and Answering) test case, we construct a set of valid reference substrings. A model’s textual response is marked as correct if it exactly matches any entry in this reference set. In evaluating spoken responses in the dialectal chit-chat test set (chitchat-dialect), we employ a supervised dialect classification model trained on multiple Chinese dialects. Each response is automatically classified, and receives a score of 1 if recognized as the target dialect, or 0 otherwise. The overall dialectal performance is then computed as the average accuracy across all test instances.

\begin{figure}[htbp]
    \centering
    \begin{subfigure}[b]{0.48\textwidth}
        \centering
        \includegraphics[width=\textwidth]{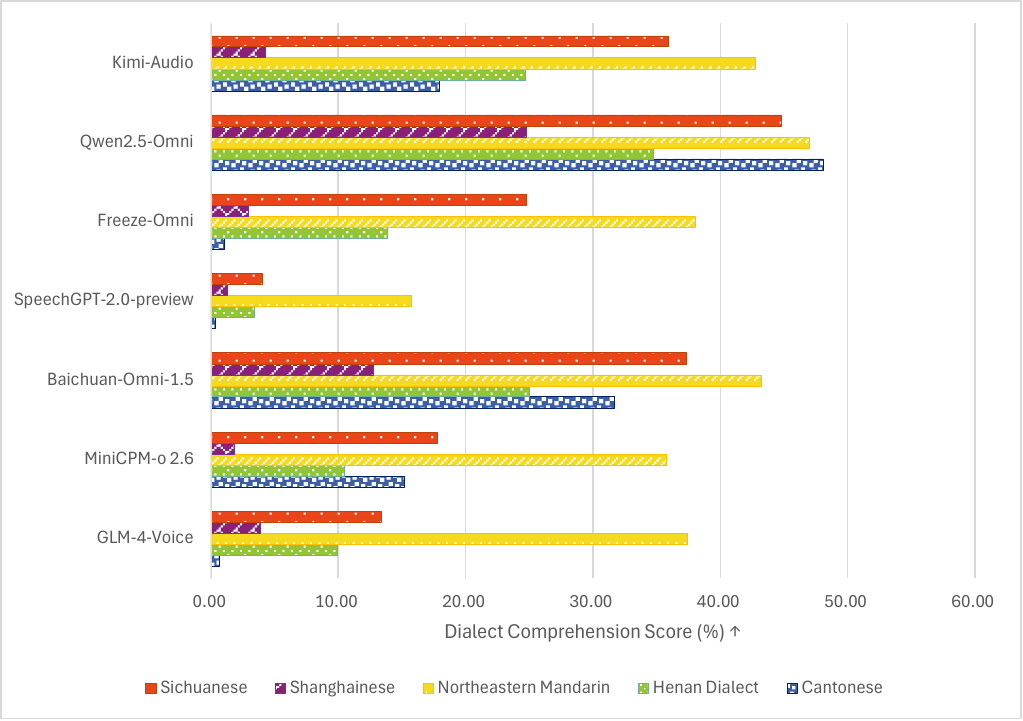}  
        \caption{Dialect Comprehension}
        \label{fig:sub1}
    \end{subfigure}
    \hspace{0.01\textwidth}
    \begin{subfigure}[b]{0.48\textwidth}
        \centering
        \includegraphics[width=\textwidth]{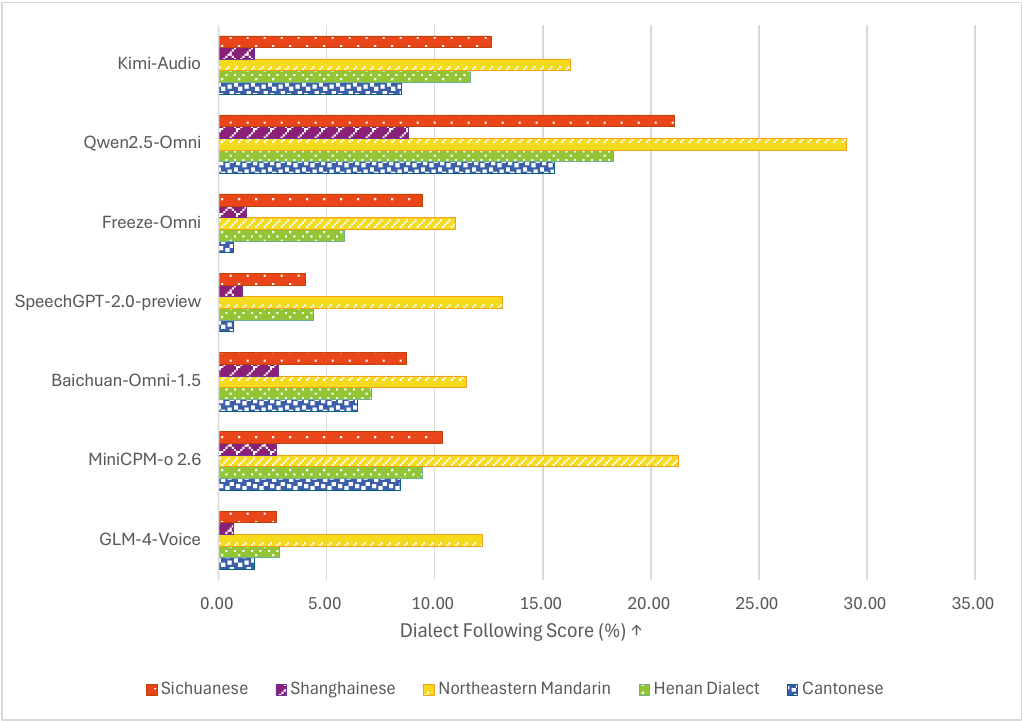}  
        \caption{Dialect Following}
        \label{fig:sub2}
    \end{subfigure}
    \caption{Results of Chinese Dialect Comprehension and Following}
    \label{fig:main}
\end{figure}

Finally, the chosen audio samples are fed to the SLMs to get their prediction results. The results in Figure~\ref{fig:sub1} and Figure~\ref{fig:sub2} show that although some models such as Baichuan-Omni-1.5 and Qwen2.5-Omni performed well in dialect comprehension, all models performed poorly in dialect generation the set of dialect chitchat.

To evaluate dialectal speech generation, we selecte our previous work, GOAT-TTS~\cite{song2025goat}, as the primary evaluation target. On the one hand, most existing speech dialogue models lack this capability; on the other hand, due to the design of the dual-modality head, GOAT-TTS not only performs TTS tasks but also possesses the ability to execute speech dialogue tasks. As shown in the table ~\ref{dialect_speech_response_results}, the Zipformer~\cite{yao2023zipformer} model trained with dialect data generated by GOAT-TTS achieves a reduction of over 70\% in WER for Cantonese, Shanghainese, and Sichuanese. Moreover, compared to the Zipformer model trained with ground-truth dialect data, it demonstrates comparable performance in Cantonese, Northeastern Mandarin, Henan dialect, and Shanghainese. This indicates that the model can efficiently extract dialect cues from speech prompts and generate authentic target dialects.

\begin{table}[ht]
    \centering
    \caption{Results for Dialectal Speech Generation}
    \label{dialect_speech_response_results}
    \small  
    \begin{tabular}{lccccc}
        \toprule
        \multirow{2}{*}{\textbf{Method}} & 
        \multicolumn{5}{c}{\textbf{ASR Word Error Rate (\%) $\downarrow$}} \\
        \cmidrule(r){2-6}
        & \textbf{Cantonese} & \textbf{Northeastern} 
        & \textbf{Henan} & \textbf{Shanghainese} & \textbf{Sichuanese} \\
        \midrule
        \textbf{Zipformer} & 89.54 & 13.63 & 61.91 & 76.81 & 41.53 \\
        \midrule
        \textbf{+ GOAT-TTS} & 26.48 & 6.52 & 24.19 & 17.34 & 14.36 \\
        \textbf{+ GT} & 23.83 & 5.55 & 21.16 & 15.60 & 11.50 \\
        \bottomrule
    \end{tabular}
\end{table}

\subsection{Context Memory}

Multi-turn conversation requires not only immediate understanding but also memory of prior turns. We evaluate the system's ability to track and use the history of dialogue, reflecting its grasp of contextual dynamics, particularly the historical interaction component, crucial to maintaining coherence and relevance in human-machine communication.

\subsubsection{Dataset}
In order to measure the model's ability to memorize the historical information in a multi-round conversation, we constructed a multi-round conversation test set with round 2\&3\&4. For multi-turn data, the user’s query in the final turn is formulated from the user’s perspective, referencing information provided in previous turns. The expected model response is a Mandarin text answer, and accuracy—measured against the provided reference answer is used as the evaluation metric. We construct 50 samples for each of the two-turn, three-turn, and four-turn settings. For each sample, a user audio input is synthesized using a speech synthesis system, with the speaker randomly selected from a pool of 80 prompt speakers. 

\subsubsection{Results}

The result in Table~\ref{multiturn memory} shows that most of these models perform relatively well on memory of multi-turn context, suggesting most of the SLMs can extract key information from the dialogue history. This observation indicates that contextual signals can be further explored and integrated in future work to enhance information association and improve the interpretive capabilities of spoken language systems.

\begin{table}[!htbp]
\caption{Results of Multi-turn Memory}
\label{multiturn memory}
\centering
\setlength{\tabcolsep}{10pt} 
\renewcommand{\arraystretch}{1.4}
\begin{tabular}{cc}
\hline
\textbf{Model}        & \textbf{Accuracy (\%) $\uparrow$} \\
\hline
GLM-4-Voice           & 80.00                     \\
MiniCPM-o 2.6         & 86.67                     \\
Baichuan-Omni-1.5     & 78.67                     \\
SpeechGPT-2.0-preview & 20.00                     \\
Freeze-Omni           & 62.67                     \\
Qwen2.5-Omni          & 88.67 \\
\hline
\end{tabular}
\end{table}

\subsection{Emotion Perception and Response}


Human speech often conveys affective intent through tone, pitch, and rhythm. This task tests the system’s ability to perceive and respond to emotional expressions, which directly corresponds to the affective cues dimension, including vocal characteristics and emotional expression, thus validating whether the model can move beyond literal understanding to empathetic interaction.

Human speech in emotional states exhibits systematic changes in physiological and acoustic correlates, such as increased muscle tension, altered respiratory patterns, fluctuations in fundamental frequency (F0), variation in intensity, and nonlinear vocal phenomena. These features are inherently tied to the speaker's affective and physiological condition, and are difficult to replicate accurately in TTS-generated speech~\cite{skerry2018towards,cowie2001emotion}. To avoid these issues, we use human-produced emotional speech for evaluation. This allows us to ensure richer and more authentic emotional expressions that better reflect the complexity of real-world emotional communication.

\subsubsection{Dataset}

We construct our emotion perception and response evaluation set based on the open-source Emotional-Speech-Dataset (ESD) \footnote{\url{https://github.com/HLTSingapore/Emotional-Speech-Data}}. This dataset contains 350 parallel utterances produced by 10 native Mandarin speakers and 10 English speakers, each expressing five emotional states: neutral, happy, angry, sad, and surprise. 
For each states, we manually select 50 audios of Mandarin to ensure that their emotions are natural and contextual, resulting in ESD-zh test set. 
Note that most of the emotional information in the test data will not be reflected in the semantics, but will only appear at the acoustic level, thus allowing for the testing of affective cues. 

\subsubsection{Results}

We evaluate the model's responses from both textual and acoustic perspectives. For textual responses, we use GPT-4o (2024-08-06) to assess the model’s understanding of the emotions conveyed in the input audio and the appropriateness of its emotional replies, as well as the degree of human-likeness in its responses. The results shown in Table~\ref{emotion result} demonstrate that only a few models can produce appropriate responses based on the user's speech input with emotions, and most of the models can only recognize the user's emotions, or even fail to do so accurately.

\begin{table}[!htbp]
    \caption{Results of Emotion Perception and Response}
    \label{emotion result}
    \centering
    \setlength{\tabcolsep}{10pt} 
    \renewcommand{\arraystretch}{1.4}
    \begin{tabular}{cc}
    \hline
    \textbf{Model}        & \multicolumn{1}{l}{\textbf{Score (\%) $\uparrow$}} \\ \hline
    GLM-4-Voice           & 35.55                                   \\
    MiniCPM-o 2.6         & 44.03                                   \\
    Baichuan-Omni-1.5     & 13.55                                   \\
    SpeechGPT-2.0-preview & 22.59                                   \\
    Freeze-Omni           & 20.72                                   \\
    Qwen2.5-Omni          & 44.83                                   \\
    Kimi-Audio            & 53.17           \\ \hline
    \end{tabular}
\end{table}

At the audio response level, we evaluate the SLM’s audio outputs by scoring them with the Emotion2Vec model\footnote{\url{https://huggingface.co/emotion2vec/emotion2vec_plus_large}}, using manually annotated emotion labels as the reference.. The results in the Table~\ref{emotion_speech_response_results} show that although some models (such as Qwen2.5-Omni and Kimi-Audio) performed well, there is still significant room for improvement in other models.

\begin{table}[!htbp]
    \caption{Results of Emotion Spoken Response Generation}
    \label{emotion_speech_response_results}
    \centering
    \setlength{\tabcolsep}{10pt} 
    \renewcommand{\arraystretch}{1.4}
    \begin{tabular}{cc}
    \hline
    \textbf{Model}        & \multicolumn{1}{l}{\textbf{Score (\%) $\uparrow$}} \\ \hline
    GLM-4-Voice           & 31.66                                   \\
    MiniCPM-o 2.6         & 34.26                                   \\
    Baichuan-Omni-1.5     & 24.74                                   \\
    SpeechGPT-2.0-preview & 27.48                                   \\
    Freeze-Omni           & 41.05                                   \\
    Qwen2.5-Omni          & 52.59                                   \\
    Kimi-Audio            & 45.48           \\ \hline
    \end{tabular}
\end{table}

\subsection{Age Perception and Response}
By inferring the speaker’s age from vocal features and adjusting the response tone accordingly, this task probes the system’s ability to integrate affective cues (like pitch and speaking rate) with implicit semantics (e.g., identity inference). It reflects the machine’s ability to personalize interaction based on latent speaker traits not explicitly stated.

\subsubsection{Dataset}
Age perception is one aspect of paralinguistic information. In an end-to-end dialogue system, the model is expected to generate appropriate responses based on the speaker's age. For different age groups (e.g., children vs. adults), the model should produce distinct responses even when the input text remains the same. To this end, we construct 150 human-behavior-based test samples to evaluate the model's ability to perceive and respond to age-related cues. Each sample avoids user input that is exclusively appropriate for a single age group (e.g., texts like “I played with my classmates in kindergarten,” which are suitable only for children). Instead, all inputs are designed to be at least plausible for adult speakers, while ensuring that human responses differ meaningfully between adults and children or elderly speakers. 

\subsubsection{Results}

Due to the open-ended nature of the Paralinguistic and Implicit Semantics tasks, conventional accuracy-based metrics are insufficient to reflect the nuanced performance of models. To address this, we adopt GPT-4o (2024-08-06) as a scoring agent and construct carefully designed evaluation prompts tailored to each sub-task. These prompts instruct the judge model to assess how well each response aligns with the paralinguistic goal, using a discrete score from 0 to 5. 

For example, in the Age Perception \& Response task, the evaluation prompt provides the model with the user’s utterance, the user’s age group (e.g., child, middle-aged, elderly), a neutral reference answer, and a stylized reference tailored to the given age group. The model is then asked to evaluate whether the generated response exhibits appropriate adaptation in tone, vocabulary, and communicative style. The scoring criteria range from 5 (highly age-appropriate and empathetic response) to 0 (inappropriate or unnatural style, lacking age awareness).

To ensure scoring reliability, each response is evaluated independently three times. We further apply a power scaling function to the average score of each sample, which increases reward sensitivity to higher scores and reduces compression effects near the top end of the scale. The final task score is computed using Equation~\ref{eq:llm_score}, where $s$ represents the raw score (0–5) and $p$ is the scaling factor.

\begin{equation}
\label{eq:llm_score}
\text{Avg}(S) = \frac{100}{|S|} \sum_{s \in S} \left(\frac{s}{5}\right)^p.
\end{equation}

Table~\ref{age result} presents the results for the Age Perception \& Response task. The findings suggest that current speech language models (SLMs) consistently underperform in this dimension. Across all evaluated models, responses show limited awareness of the speaker’s age group, often defaulting to a generic or adult-oriented reply style. Even in cases where the model successfully inferred the age (e.g., recognizing the speaker as a child or elderly), it rarely adapted its language tone, vocabulary, or sentence structure accordingly.

\begin{table}[!htbp]
\caption{Results of Age Perception and Response}
\label{age result}
\centering
\setlength{\tabcolsep}{10pt} 
\renewcommand{\arraystretch}{1.4}
\begin{tabular}{cc}
\hline
\textbf{Model}        & \textbf{Score (\%) $\uparrow$} \\
\hline
GLM-4-Voice           & 27.81               \\
MiniCPM-o 2.6         & 34.56               \\
Baichuan-Omni-1.5     & 12.24               \\
SpeechGPT-2.0-preview & 23.63               \\
Freeze-Omni           & 13.68               \\
Qwen2.5-Omni          & 42.51               \\
Kimi-Audio            & 22.77        \\
\hline
\end{tabular}
\end{table}

\subsection{Non-verbal Information}

Human communication includes rich non-verbal signals such as sighs, coughs, and pauses, which often imply intent, emotion, or conversational dynamics. This evaluation explores how the model handles these cues, engaging with affective cues (non-verbal sounds), contextual dynamics (e.g., pauses, background sounds), and even implicit semantics (e.g., inferred emotional or social intent).

\subsubsection{Dataset}
Non-verbal information is one aspect of paralinguistic information. In an interaction scene, non-verbal vocal signals often convey information beyond semantics. For example, a speaker clearing their throat during a conversation may indicate dissatisfaction—an intent that cannot be captured purely at the semantic level. To evaluate the model's responsiveness to such non-verbal vocal signals, we use four types of real non-speech sounds, coughing, throat clearing, laughter, and sneezing with 3,500 authentic audio segments for each category. These segments are randomly sampled and concatenated either before or after the original user input in the existing AQA test set. The expected model response includes both an appropriate answer to the user's input and an expression of attentiveness or concern in response to the non-verbal vocal signal.

\subsubsection{Results}
The result is shown in Table~\ref{non-verbal results}, these models exhibit limited capability in handling non-verbal vocal signals. 
While Kimi-Audio exhibits a certain level of recognition capability, this behavior aligns more closely with audio event detection (AED) tasks rather than genuine perception and response to non-verbal cues.

\begin{table}[!htbp]
\caption{Results of Responsiveness of Models to Non-verbal Vocal Signals}
\label{non-verbal results}
\centering
\setlength{\tabcolsep}{10pt} 
\renewcommand{\arraystretch}{1.4}
\begin{tabular}{cc}
\hline
\textbf{Model}        & \textbf{Score (\%) $\uparrow$} \\
\hline
GLM-4-Voice           & 1.89                \\
MiniCPM-o 2.6         & 2.08                \\
Baichuan-Omni-1.5     & 1.80                \\
SpeechGPT-2.0-preview & 1.52                \\
Freeze-Omni           & 1.85                \\
Qwen2.5-Omni          & 2.19                \\
Kimi-Audio            & 9.19                \\
\hline
\end{tabular}
\end{table}

\section{Discussion}

These findings indicate that existing SLMs lack the ability to fully consider and integrate the nuances of BoSS-related attributes. As a result, their outputs often fail to deliver the empathetic, contextually aware, and emotionally intelligent interactions expected in natural human communication. This limitation highlights the need for further exploration and refinement of BoSS dimensions.

Moreover, the performance gaps observed across different BoSS dimensions, such as dialect comprehension, age-aware responses, and nonverbal cue interpretation, suggesting that current models still treat speech predominantly as a sequence of textual tokens, overlooking the rich vocal and contextual signals that shape human intention and nuance. These deficiencies are particularly critical for real-world applications, such as healthcare, education, and customer service, where emotional appropriateness and social awareness are essential.

To move forward, it is necessary to develop models and evaluation frameworks that explicitly incorporate multidimensional beyond-semantic cues. This includes learning from paralinguistic patterns, integrating long-range context memory, modeling emotional dynamics, and enabling adaptive, socially sensitive responses. BoSS offers a conceptual and technical foundation for this pursuit, and advancing it will be essential for building next-generation spoken interaction systems that can truly understand, which means not just transcribe human communication.

\section{Conclusion}
This paper introduces and formalizes Beyond-Semantic Speech (BoSS), aiming to bridge the gap between explicit linguistic processing and the rich, implicit layers of human speech communication. BoSS highlights the importance of multidimensional features—such as affective cues, contextual dynamics, and implicit semantics—that shape how meaning is conveyed and interpreted in real-world interactions. We propose a computational framework grounded in Relevance Theory and probabilistic modeling to represent and infer the most plausible meaning hypotheses based on cognitive effects and processing effort. 
Through preliminary evaluations in five BoSS-related dimensions, including dialect comprehension, contextual memory, emotional recognition, age perception, and non-verbal interpretation, we find that current spoken language models are still not able to capture these beyond-semantic aspects. This gap emphasizes the need for further research into integrating fine-grained vocal and contextual signals into speech processing systems.

\bibliography{sn-bibliography}

\end{document}